# Spin dynamics of electrons in the first excited subband of a high-mobility low-density 2D electron system

Haihui Luo, Xuan Qian, Xuezhong Ruan, Yang Ji[*]

*SKLSM, Institute of Semiconductors, Chinese Academy of Sciences,*

*Beijing 100083, People's Republic of China*

V. Umansky

*Braun Center for Submicron Research, Department of Condensed Matter Physics,*

*Weizmann Institute of Science, Rehovot 76100, Israel*

[*]Author to whom correspondence should be addressed, jiyang@red.semi.ac.cn

We report on time-resolved Kerr rotation measurements of spin coherence of electrons in the first excited subband of a high-mobility low-density two-dimensional electron system in a GaAs/Al$_{0.35}$Ga$_{0.65}$As heterostructure. While the transverse spin lifetime ($T_2^*$) of electrons decreases monotonically with increasing magnetic field, it has a non-monotonic dependence on the temperature, with a peak value of 596 ps at 36 K, indicating the effect of inter-subband electron-electron scattering on the electron spin relaxation. The spin lifetime may be long enough for potential device application with electrons in excited subbands.

Electron spin manipulation is among the hottest topics in semiconductor spintronics research, which aims to realize next-generation devices via utilizing the spin degree of freedom of electrons.[1-3] Due to its important role in achieving effective spin manipulation, the spin dynamics of carriers in semiconductors has been extensively studied in last decades via many kinds of spin-sensitive spectroscopy techniques,[4-7] among which is the powerful time-resolved Kerr rotation (TRKR) measurement, being widely used to experimentally study spin dynamics in bulk materials[8,9] and quantum-confined system such as quantum wells[10,11] and a single quantum dot.[12,13]

Most of the experiments to date are concerned with spin coherent processes of electrons in ground state. However, there is not much research on the spin dynamics of electrons in excited states of quantum-confined system so far.[14,15] Here we report on experimental investigation of the spin coherence of electrons in the first excited subbband of a two-dimensional electron system (2DES) with high mobility and low density. With TRKR techniques, we measured both the magnetic field dependence (0–4 T) and the temperature dependence (1.5–80 K) of the transverse spin lifetime ($T_2^*$) of electrons. While $T_2^*$ decreases monotonically with an increasing magnetic field, it has a non-monotonic dependence on the temperature, with a peak value of 596 ps at 36 K (measured at magnetic field $B$=1.5 T), indicating the effect of electron-electron scattering on the electron spin decoherence[4,16,17] while in this case inter-subband scattering is dominant over intra-subband scattering. The spin lifetime may be long enough for potential device application with electrons in excited subbands.

The sample is a modulation-doped $GaAs/Al_{0.35}Ga_{0.65}As$ 2DES sample, with a mobility of $\mu$=3.2×10$^6$ cm$^2$/Vs and an electron density of $n_{2D}$=9.6×10$^{10}$ cm$^{-2}$ at $T$=4.2 K.[17] The 2DES sample was mounted inside an optical cryostat with a tunable temperature from 1.5 K to 300 K and a transverse magnetic field up to 10 T provided by a superconducting split coil. Photoluminescence (PL) spectrum measurement was performed with a He-Ne laser at temperature $T$=2 K and magnetic field $B$=0 T to determine the energy levels of the 2DES.

Figure 1(a) shows the PL spectrum at 2 K, which has two peaks at $E$=1.519 ev and $E$=1.578 ev. The PL peak at $E$=1.519 ev is attributed to the recombination between holes and electrons in the ground subband (GS) of the 2DES. Note that the lower energy part of this PL peak may partly come from the recombination of excitons in the bulk GaAs buffer layer. The peak at $E$=1.578 ev is attributed to the recombination of holes and electrons in the first excited subband (FES) of the 2DES. Accordingly, we may draw the energy levels of the 2DES schematically in Fig. 1(b).

After determining the energy levels of the 2DES, we performed TRKR measurements to study the dynamical processes of spin-polarized electrons in the FES of the sample. In the TRKR measurements, the sample was excited near normal incidence with degenerate pump and probe beams from a mode-locked Ti: sapphire pulsed laser, which has a pulse width of 150 fs, a repetition rate of 76 MHz, and a tunable wavelength between 700 nm and 980 nm. The pump and probe beams were focused to a spot of ~100 μm in diameter. The circular polarization of the pump beam was modulated with a photoelastic modulator at 50 kHz for lock-in detection. With the magnetic field being perpendicular to the laser beams (Voigt geometry), the Kerr rotation $\theta(\Delta t)$ of the linearly polarized probe light pulse after a time delay $\Delta t$ measures the projection of the net spin polarization along the direction of the light as it precesses about the direction of the magnetic field. When we studied the spin dynamics of electrons in the FES, we did the TRKR measurement with photon energy $E_p$=1.578 ev and pump/probe power of 10 mW/1 mW.[18]

Figure 2(a) shows two TRKR traces measured at $T$=1.5 K and $B$=2.5 T with $E_p$ =1.519 ev and $E_p$ =1.578 ev, respectively. Both of them have beating features, as indicated by the two peaks in the fast Fourier transform (FFT) spectra shown in the inset of the Fig. 2(a). The low frequency and high frequency parts in the $E_p$ =1.519 ev trace are attributed to electrons in the GS and the bulk GaAs, respectively, as in our previous work.[17] As for the $E_p$ =1.578 ev case, the high frequency part is ascribed to the spin precession of the electrons in the FES of the 2DES, while the low frequency part is attributed to the photoexcited free holes.[19] The Larmor precession frequency of the electrons in the FES is slightly lower than in the GS because the wave function of the electrons in the FES penetrates more into the potential barrier $Al_{0.35}Ga_{0.65}As$ than that of the GS. The g factor of GaAs and $Al_{0.35}Ga_{0.65}As$ are −0.44 and +0.5 respectively,[20] the more spread of electron wave function in the $Al_{0.35}Ga_{0.65}As$ barrier leads to a smaller electron g factor (in absolute value).

The FES TRKR trace at $E_p$ =1.578 ev is redrawn in the top part of Fig. 2(b). The Kerr signal coming from electrons in the FES of the 2DES and the photoexcited holes can be decomposed via their distinct Larmor precession frequencies. We fit the experimental data with the following formula:

$$\theta_k(\Delta t) = A_e \exp(-\frac{\Delta t}{T_{2,e}^*})\cos(2\pi v_e \Delta t + \phi_e) + A_h \exp(-\frac{\Delta t}{T_{2,h}^*})\cos(2\pi v_h \Delta t + \phi_h) \quad (1)$$

where $A_e(A_h)$ is the electron (hole) signal amplitude at pump-probe delay $\Delta t$=0, $T_{2,e}^*$ ($T_{2,h}^*$) is the electron (hole) spin lifetime, $v_e(v_h)$ is the Larmor precession frequency of the

electron(hole), and $\phi_e$ ($\phi_h$) is a phase shift. The contributions from electrons in the FES and holes are shown respectively at the middle and bottom of Fig. 2(b), respectively, the sum of which fits well to the experimental data, as shown by the solid line in the upper panel. Accordingly, we obtained the transverse spin lifetime $T_2^*$ and the Larmor precession frequencies $\nu$ for both of the low and high frequency spin precession. The g factors of electrons in the FES and holes can be thus deduced by $\nu=g\mu_B B/h$, where $\mu_B$ is the Bohr magneton, $B$ is the transverse magnetic field, and $h$ is the Plank's constant. A spin lifetime of 355 ps (close to the result of other 2DES sample[21]) and a g factor of 0.405 can be obtained for the electrons in the FES, while a spin lifetime of 49 ps and a g factor of 0.106 for the holes. In the following, we will only focus on the FES part.

    Figure 3 shows TRKR signals at different magnetic field $B$ at $T$=1.5 K. Fitting them with Eq. (1), we obtain the dependence of the $T_2^*$ of the electrons in the FES on the magnetic field, as shown in the inset. A long-lived electron spin coherence with $T_2^*$=577 ps is found at $B$=0.5 T. With increasing $B$ up to 4 T, it decreases monotonically to 290 ps.

    Figure 4(a) shows TRKR traces at various temperatures at $B$=1.5 T.[22] As can be seen in this figure, with increasing temperature, the decay of the envelope of the oscillating Kerr signal becomes more slowly from 1.5 K to 36 K, and slightly faster since then (from 36K to 80K). In order to figure out the temperature dependence of spin lifetime of the electrons in the FES, we measured the TRKR trace from 1.5 K to 80 K in detail, and the spin lifetimes are shown in Fig. 4(b). A peak value of 596 ps is seen at 36 K in the electron spin lifetime $T_2^*$. The maximum is superimposed on an increasing spin lifetime background from 410 ps at 1.5 K to 507 ps at 80 K. The temperature dependence of the $T_2^*$ of the electrons in the

FES is similar to that in the Gs of the 2DES.[17] However, the maximal $T_2^*$ of the electrons in the FES appears at 36 K, while that of the electrons in the GS appears at 14 K (Ref. 17), the reason for the difference goes as the following.

There are two main contributions for the transverse spin lifetime of the electrons in the FES. One is the relaxation process in which the electrons in the FES relax into the GS or recombine with holes: this is an annihilation process of the electron in the FES which determines the electron relaxation time $T_1$; the other is the phase smearing of electrons which remain in the FES: it determines the "pure" electron spin dephasing time $T_{22}^*$. This may be expressed by the following formula:

$$\frac{1}{T_2^*} = \frac{1}{T_1} + \frac{1}{T_{22}^*} \qquad (2)$$

$T_1$ is mainly determined by electron-phonon scattering, electron-impurity scattering and electron-hole recombination. The high mobility of the 2DES sample used here ensures that the electron-impurity scattering is weak. The phonon assisted relaxation is also weak, since it is difficult for the LO and TO phonons to fulfill the conservation laws of both the momentum and the energy (the phonon bottle-neck effect in a quantum confined system[23]). The electron–hole recombination is also weak since optically generated holes are swept quickly into the GaAs buffer layer. A long $T_1$ can thus be expected: it could be even longer than 1 ns and does not depend much on the temperature in the regime below 100 K.[24] Thus, the main contribution to the $T_2^*$ dependence on the temperature (and/or the magnetic field) comes from $T_{22}^*$.

The field dependence of $T_2^*$ may come from the low density feature of the sample. Since the electron density is very low, there may be puddles of electrons with different density, leading to an inhomogeneous g factor (like the case in localized two dimensional holes and the electrons in quantum dots[10,25]) and a spread of Larmor precession frequencies in a transverse magnetic field: $\Delta \nu = \Delta g \mu_B B/h$, where $\Delta g$ is the inhomogeneity of the electron g factor. Note that with a fixed $\Delta g$, a higher field lead to a larger phase dispersion, and further a faster spin dephasing.[5] The relationship can be described by $1/T_{22}^* = 1/T_{22}^*(0) + \Delta g \mu_B B/\sqrt{2}\hbar$,[25] where $T_{22}^*(0)$ is the zero-field "pure" spin dephasing time. Therefore, Eq. (2) can be expressed as:

$$\frac{1}{T_2^*} = \frac{1}{T_1} + \frac{1}{T_{22}^*(0)} + \frac{\Delta g \mu_B B}{\sqrt{2}\hbar} \tag{3}$$

The solid line in Fig. 3(b) shows a $1/B$ fit to the $T_2^*$ data, from which a $T_1$=1698 ps and a $T_{22}^*(0)$=1100 ps is obtained.

As the D'yakonov-Perel' (DP) mechanism says, the spin relaxation rate is determined by $\tau^{-1} = \langle \Omega(\mathbf{K})^2 \rangle \tau_p(\mathbf{k})$, where $\tau_p(\mathbf{k})$ is the momentum relaxation time.[5] The electron-phonon scattering has a temperature dependence of $\tau_{eac} \propto T^{-3/2}$.[26] An increasing temperature leads to a stronger momentum scattering, in other word, a shorter momentum scattering time $\tau_p$. This, in turn, induces an increasing $T_2^*$ via DP mechanism. Consequently, there is an increasing $T_2^*$ background with rising temperature. In a high-mobility low-density 2DEG, electron-electron Coulomb scattering dominates $\tau_p$ at low temperature, contributing to the $T_2^*$ peak at 36 K shown in Fig. 4(b), with similar arguments in previous theory and experimental work.[4,16,17] However, the peak is significantly shifted toward higher temperature, in contrast with the similar peak in our previous work on the electrons of the GS.[17] Such a difference may be due to the difference between the electron-electron scattering mechanism in these two cases. If the electrons are excited to the GS, the intra-subband electron-electron scattering is dominant. However, when the electrons are excited to the FES, because of the low electron density on the FES, the electron scattering are dominated by inter-subband scattering, and the items of electron distribution function appears in the $1/\tau_{e\text{-}e}$ expression would be dramatically different, inducing the different temperature dependence of $1/\tau_{e\text{-}e}$.[27]

In summary, we have experimentally studied the spin dynamics of electrons in the first excited subband in a high-mobility low-density 2DES. While $T_2^*$ decreases monotonically with an increasing magnetic field at 1.5 K, it has a non-monotonic dependence on the temperature in a magnetic field of 1.5 T, with a peak value of 596 ps at 36 K, indicating the effect of inter-subband electron-electron scattering on the electron spin decoherence. The spin lifetime may be long enough for potential device application with electrons in excited subbands.

We thank M.W. Wu and J.H. Jiang for fruitful discussions. This work was supported by the NSFC under Grants No. 10425419, National Basic Research Program of China (No. 2007CB924900, No. 2009CB929301) and the Knowledge Innovation Project of Chinese Academy of Sciences.

Shabaev, A.L. Efros, I.A. Merkulov, V. Stavarache, D. Reuter and A. Wieck, *ibid.* **96**, 227401 (2006).

26. J. Bardeen and W. Shockley, Phys. Rev. **80**, 72 (1950).

27. B.Y.K. Hu and K. Flensberg, Phys. Rev. B **53**, 10072 (1996 ).

**Figure captions**

FIG. 1. (a) PL spectrum at $T$=2 K and $B$=0 T . (b) Schematical energy levels of the 2DES (not to scale), the two broken line with arrows indicate the excitation with $E$ =1.519 ev and $E$ =1.578 ev, respectively.

FIG. 2. (Color online) (a) TRKR traces were measured with $E_p$ =1.519 ev and $E_p$ =1.578 ev at $T$=1.5 K and $B$=2.5 T. Inset: FFTs of the corresponding TRKR traces. (b) Top: the experimental TRKR trace with $E_p$ =1.578 ev at $T$=1.5 K and $B$=2.5 T and the fitting curve (the solid line). Middle: the extracted TRKR signal of the electrons in the FES. Bottom: the extracted TRKR signal of the holes.

FIG. 3. (Color online) TRKR traces at different magnetic field ($B$) at $T$=1.5 K. Inset: The transverse spin lifetime as a function of $B$ for electrons in the FES (squares) and the fitting curve (the solid line) with Eq. (3).

FIG. 4. (Color online) (a) TRKR traces at different temperatures (measured at $B$=1.5 T). (b) Spin lifetime as a function of temperature for electrons in the FES (squares) and holes (circles).

**Figure 1**

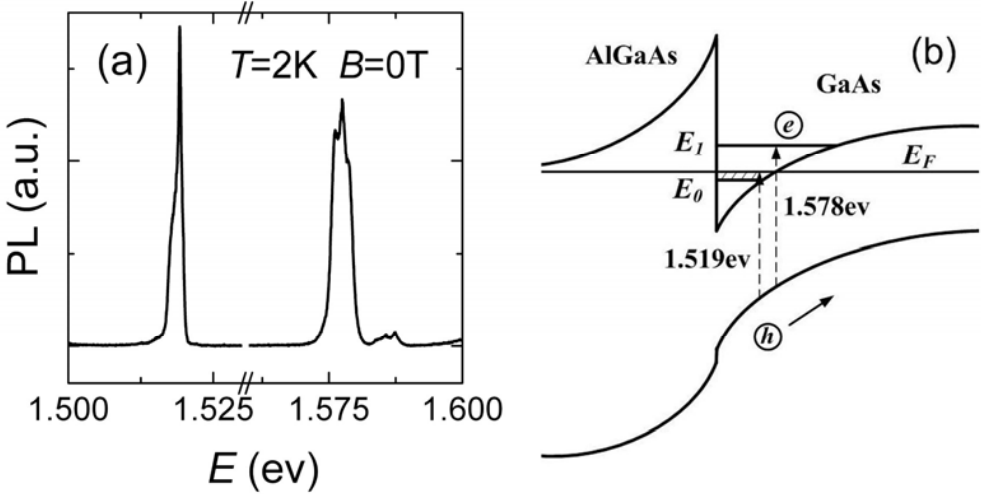

**Figure 2**

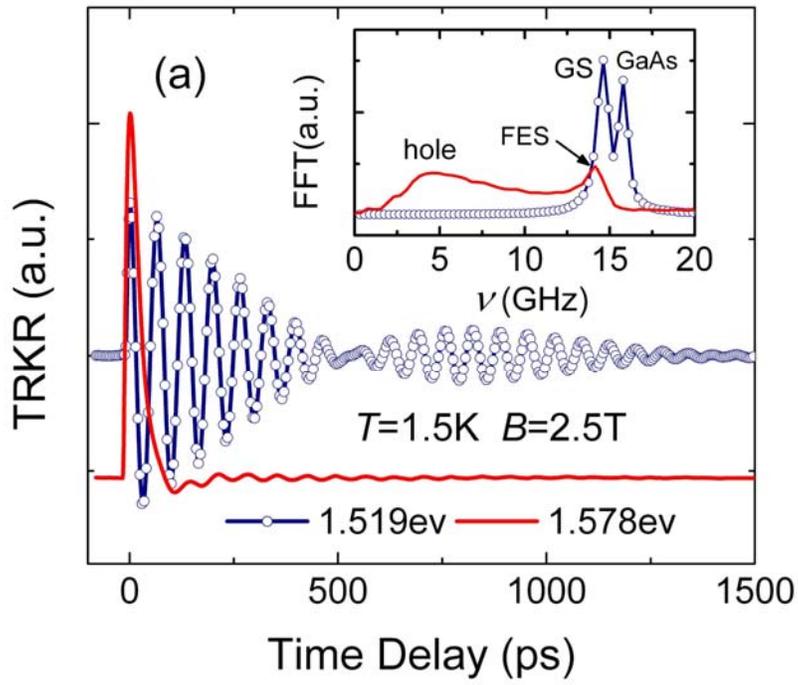

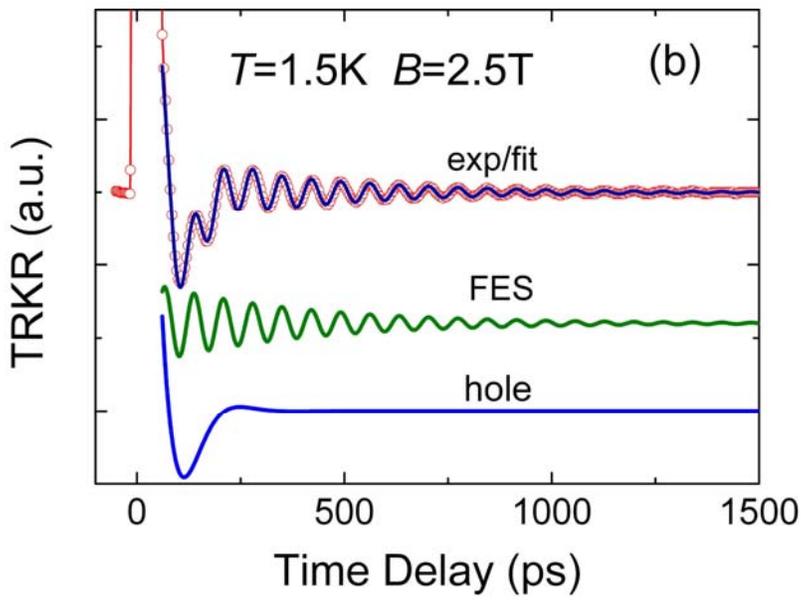

**Figure 3**

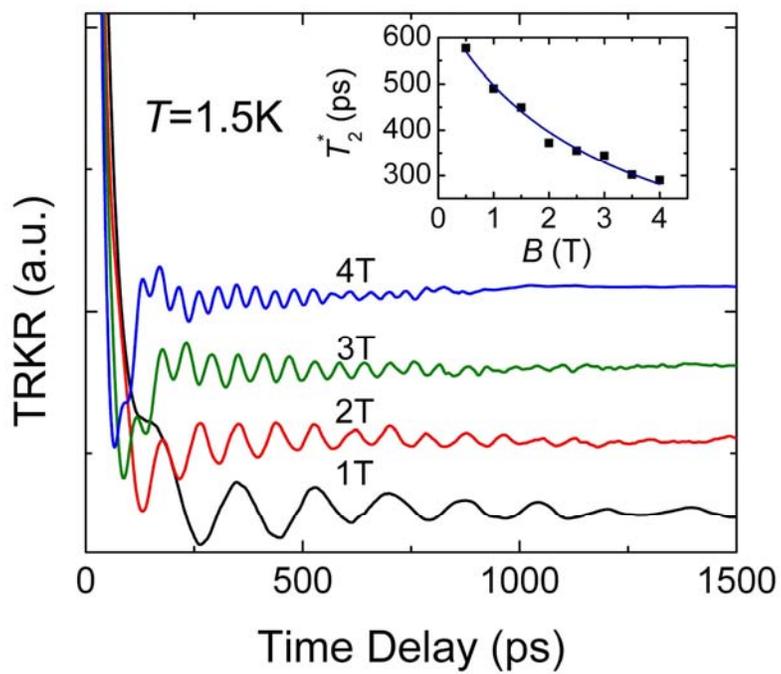

**Figure 4**

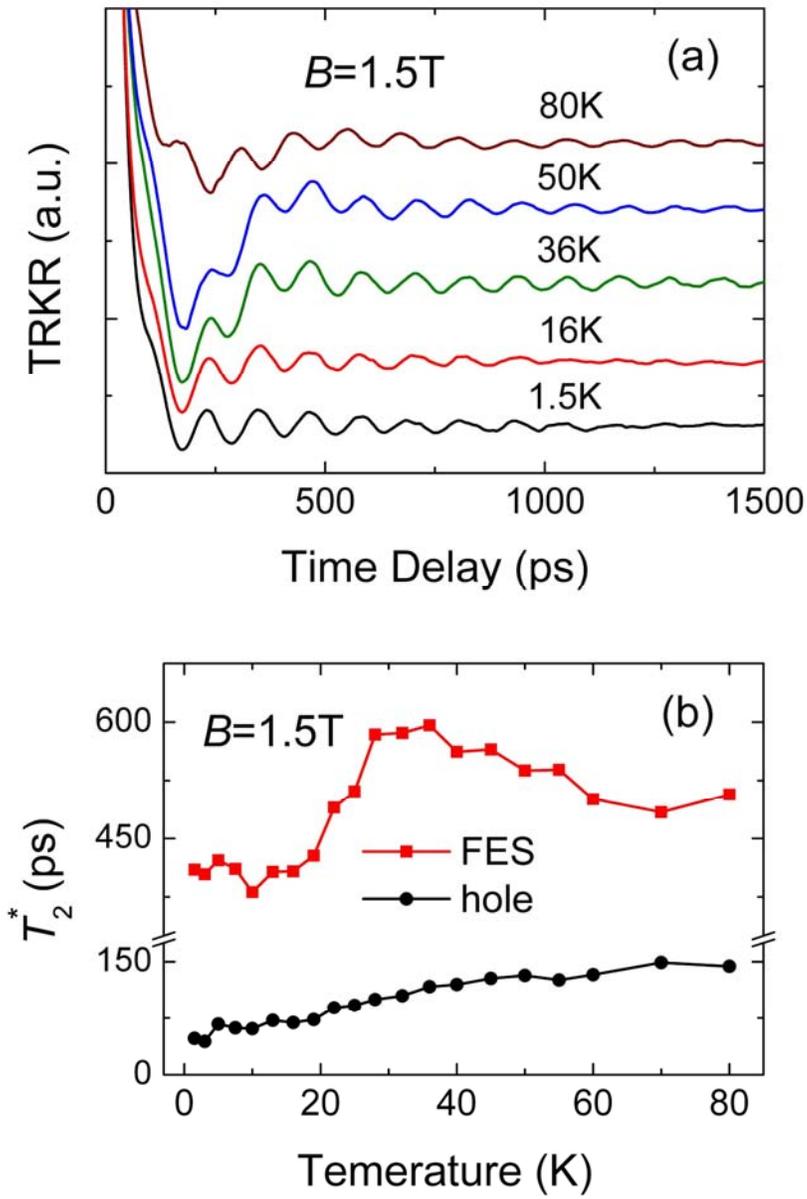